# Exploring Crossmodal Interaction of Tactile and Visual Cues on Temperature Perception in Virtual Reality: a Preliminary Study

Helfenstein-Didier, CD, Clémentine

Univ Lyon, ENISE, LTDS, UMR 5513 CNRS, 58 rue Jean Parot, 42023 Saint-Etienne cedex 2, France, clementine.didier@enise.fr

Dhouib, AD, Amira

National Engineering School (ENISE), 58 rue Jean Parot, 42023 Saint-Etienne cedex 2, France, amiradhouib@live.fr

Favre, FF, Florent

National Engineering School (ENISE), 58 rue Jean Parot, 42023 Saint-Etienne cedex 2, France

Pascal, JP, Jonathan

National Engineering School (ENISE), 58 rue Jean Parot, 42023 Saint-Etienne cedex 2, France

Baert, PB, Patrick

National Engineering School (ENISE), 58 rue Jean Parot, 42023 Saint-Etienne cedex 2, France, patrick.baert@enise.fr

Thanks to the digital revolution, virtual reality (VR) has getting popularity due to its capacity to immerse users into virtual environments (VEs). VEs are typically limited to visual and auditory cues; however, recent results show that multiple sensory modalities increase the user's immersion. In this study, an experimental protocol is proposed to recreate multiple tactile, in particular thermal, sensations in VR. The aim is twofold: (1) studying the performance of different devices for creating warm and cold sensations with regards to their efficiency and acoustic disturbance; and (2) investigating the interdependency between visual and tactile stimuli in the perception of temperature. 14 participants performed two experimental studies. Our results show no acoustic disturbance of the materials used. Spot projector is more efficient than fan heater to create a warm sensation; fan + water spray is more efficient than fan alone to create cold sensation. Moreover, no significant contribution of visual cue on the thermal perception was found except for the extremely cold simulation (snow visualization and thermal stimulation performed with fan + water spray).

## 1 INTRODUCTION

In recent years, there has been an increasing interest in Virtual Reality (VR). This interest is reflected in the use of this technology in several fields like education, industrial maintenance, healthcare, etc. [1-3]. With the growth of the use of VR, companies and researchers are not anymore focusing exclusively on visual cues to make the Virtual Environments (VE) more realistic or to increase the user's immersion of the VE. In fact, stimulating the different human senses in VR represents a possible way to achieve these objectives. Obrist et al. [4] highlight the potential of designing multisensory experience leading to the creation of new products and experiences. Researchers investigate the effect of multisensory stimulation on different parameters (quality of experience, sense of presence, immersion, memory, etc.) [5-7]. 84.8% of the studies that Melo et al. [3] analyzed show a positive impact of multisensory VR experiences. According to them, haptics is the most commonly used stimulus in multisensory VR systems (86.6%).

Haptic cues include cutaneaous cues that allow us to detect stimuli such as touch, vibrations and temperature [8]. Some researches aim at implementing thermal cues in VR. There exist various non-contact thermal feedback [6, 9]. Cold cues is less studied as fans are more generally used to simulate wind than cold [10-12]. Contact thermal stimulations are also expanding [5, 13-14]. Others are very sophisticated [7, 15-16].

Although these devices are increasingly used, they focus on the impact of multisensorial cues on parameters such as quality of experience or sense of presence [5, 7, 13]. There exist few studies that evaluate the performance of new devices for warm and cold sensations and conclusions are heterogeneous [15-16]. To the best of our knowledge, there is not yet a proposal that assess the acoustic disturbance during the use of these devices. Motivated by this lack, the main objective of this research is to better understand the impact of acoustic disturbance on thermical sensation. Finally, we compare our proposal with respect to the existing studies

## 2 METHODOLOGY

We investigate the following research questions:

Q1. What is the performance of different devices for creating warm and cold sensations with regards to their efficiency and acoustic disturbance?

Q2. Is there any interdependency between visual and tactile stimuli in the perception of temperature?

### 2.1 Participants

Fourteen volunteers (7 males, 7 females) aged from 24 to 68 years (mean=37.5±10.97), wearing a t-shirt and pants, participated to our experiments. All of them have no or a limited experience with the VE. Volunteers participated after giving informed consent and did not receive any monetary compensation.

### 2.2 Experimental Design

The study was carried out for approximately 30min per participant (from arrival to departure). Our study had 2 scenarios: scenario 1 thermic stimulation without a visual cue (black screen in the HMD), scenario 2 thermic stimulation with a visual cue (chimney for warm sensation, snowing mountains for cold sensation - available on steam VR). During each scenario, the same cold and warm stimulation are used. Different devices have been used in each test area to recreate the feeling of median/high cold, median/high warm (see Table 1). The initial room temperature (22.88±1.09°C), hygrometry (36.14±9.03 %) and level of sound (38.9±0.2dBA) were measured before each experiment. All the devices were manually activated by the experimenter.

Table 1: Characteristics of the used devices during the experiments

| Devices | | Characteristics | Settings used for experiments | Sound level (dBA) | Speed of wind m/s |
|---|---|---|---|---|---|
| Heat systems | Fan heater | - 1000 W (level 1)<br>- 2000 W (level 2) | Position of device: Pelvis height | 48.9 – 50.5 | 0 – 0.3 |
| | Spot projector | - 1200W<br>- Produces radiation | 0.5m from the participant | 39.0 | - |
| Cold systems | Electric fan | - 50 W<br>- Two speed levels | Position of device: Pelvis height 1m from the participant | 43.5 – 44.8 (level 1)<br>51.5 – 53.8 (level 2) | 0.7 – 1.6 (level 1)<br>1.5 – 3 (level 2) |
| | Electric fan + Evian sprayer | - 50 W<br>- Two speed levels<br>- Generate humidity | | 43.7 – 45.5 (level 1)<br>51.0 – 53.9 (level 2) | 0.7 – 1.6 (level 1)<br>1.5 – 3 (level 2) |

**{Cold area}.** Two set-up were tested to generate the cold stimuli: an electric fan (with two levels), and the same fan with a water spray. Level 1 corresponds to a speed of wind between 0.7 and 1.6m/s; level 2 corresponds to a speed of wind between 1.5 and 3m/s when the fan and sprayer were placed at 1m from the participants. Instantaneous level of sound was registered with a sonometer (see Table1).

**{Warm area}.** An electric heater and a spot projector are used. All the devices were placed at pelvis height of the volunteer. The power level of the heater was modified. Level 1 and level 2 correspond to a speed of blowing air between 0 and 0.3m/s when the heater was placed at 0.5m from the participants. Instantaneous level of sound was registered with a sonometer (see Table1).

After each stimulation exposure, participants were required to answer questions on thermal sensation and noise disturbance. Before starting, a thermal scale from 0 to 5 (0 refers to extremely cold, 5 refers to extremely warm) was showed to explain the scoring system. Participants answered the following questions:

How do you rate the thermal sensation? *{rate from 0 to 5}*. Did the noise bother you? *{answer yes or no}*.

## 3 RESULTS

We look into the performance of different devices for creating warm and cold sensations with regards to their efficiency and acoustic disturbance; and if there is any interdependency between visual and tactile stimuli in the perception of temperature. A Wilcoxon paired test was used to compare the sensation of cold, wind, or heat as function as used devices.

### 3.1 Noise disturbance

Sound levels during the experiment are reported in Table 1. 100% of the participants reveal to have no sound disturbance during scenario 1. Most of the participants were not bothered by the noise produced by the different devices (Table 2). The heater is noisier than other devices. 28.57% of volunteers were disturbed by its noise.

Table 2: Questionnaire results of question "Did the noise bother you?" *{answer yes or no}* in scenario 2

| | Heater level 1 | Heater level 2 | Spot projector | Fan level 1 | Fan level 2 |
|---|---|---|---|---|---|
| Yes | 28.57% | 28.57% | 0% | 0% | 7.14% |
| No | 71.43% | 71.43% | 100% | 100% | 92.86% |

### 3.2 Performance of different devices

**Cold devices.** Figure 1a illustrates the mean values and standard deviations regarding the perception of temperature in both scenarios and for each level of fan. The level of cold feeling is increasing with the intensity (level) of the fan (scoring are decreasing). For example in scenario 1, cold sensation is at 2.18±0.32 for level 1 and 1.82±0.42 for level 2 (resp. 2.07±0.27 and 1.82±0.32 in scenario 2). Both levels are statistically different ($p=0.005$ for test 1, and $p=0.01$ for test 2).

Figure 1b illustrates the mean values regarding the question "How do you rate the thermal sensation?" in both scenarios and for each level of fan+spray. The level of cold feeling is increasing with the intensity (level) of the fan, e.g. for scenario 1: 1.64±0.41 for level 1 and 1.36±0.53 for level 2 (respectively 1.36±0.41 and 1.07±0.43 in scenario 2). Both levels are statistically different ($p=0.015$ for test 1, and $p=0.025$ for test 2).

**Warm devices.** Figure 2 shows the mean values regarding question "How much do you feel heat?" for both scenarios. The warm sensation is higher with the projector than with the heater. For example, without VE (resp. with VE) means the sensation of heat is at 4.25±0.43 for spot (resp. 4.36±0.53) compared to 3.68±0.32 and 4.21±0.32 at level 1 and 2 of the heater (resp. 3.71±0.26 and 4.32±0.37). The differences are statistically different ($p=0.002$ at level 1 of the heater, for test 1 and $p=0.005$ at level 1 of the heater).

The heating sensation significantly increases for both scenarios between heater level 1 and the other devices.

No difference ($p>0.05$) was found, for both scenarios, between heater at level 2 and spot projector. Since spot projector has shown to produce no noise disturbance, which is not the case of fan heater level 2, those results thus suggest the spot projector as the recommended device for producing heat sensation.

### 3.3 Effect of visual stimuli

**Cold simulation.** As we can see in Figure 1, no significant difference was found between scenarios (with or without visual stimuli) for fan devices alone. However, the addition of spray in scenario 2 (with visual stimuli) leads to an increase of the cold sensation. Especially at level 2, a significant difference is found ($p=0.025$).

**Warm simulation.** The warm sensation is slightly increasing with visual stimuli as mean scores are higher for scenario 2 than for scenario 1. However, those results show no statistical difference ($p>0.48$) between both scenarios.

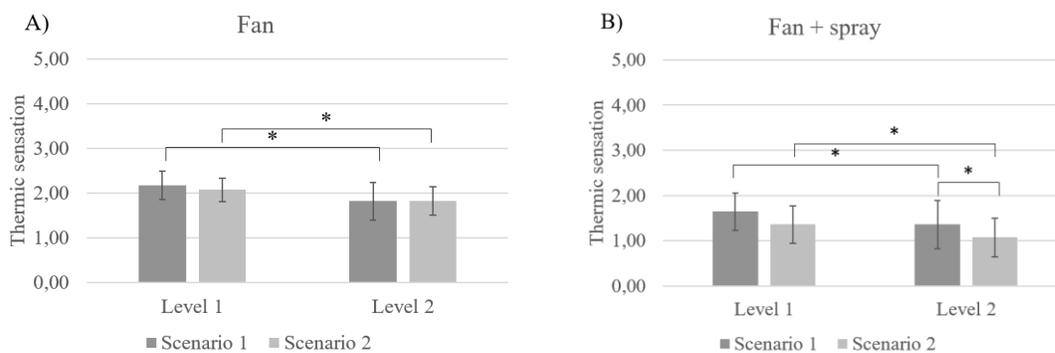

Figure 1: Perceived thermic sensation results for cold simulation, for both scenarios (n=14) with black display (scenario 1) and visual display (scenario 2). 0 represents "extremely cold", and 5 "extremely warm".

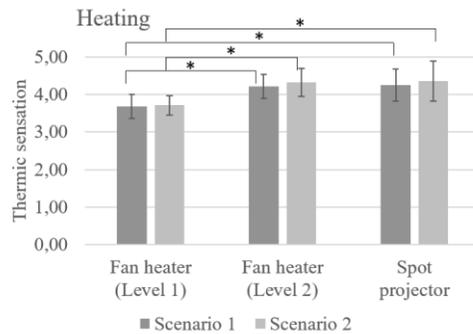

Figure 2: Perceived thermic sensation results for warm simulation, for both scenarios (n=14)

## 4 DISCUSSION

This study has enabled us to highlight notable aspects: 1) the performance of different devices for creating warm and cold sensations with regards to their efficiency and acoustic disturbance and 2) the limited interdependency between visual and tactile stimuli in the perception of temperature.

As the primary findings of our study, the way to produce cold has an impact on temperature perception. Significant results were observed: for scenario 1: p=0.003 for fan at level 1 and p=0.004 for fan at level 2; for scenario 2, p=0.001 for fan at level 1 and 2. The way to produce warm has an impact on the perceived temperature on both scenarios only when heater is at level 1 (p=0.002 for scenario 1 and p=0.005 for scenario 2). No significant difference was observed between heater at level 2 and spot projector. However, some devices are noisier and can disturb volunteers in their immersion. For heat, the projector turned out to be quieter than the heater where 28.57% of volunteers reveals to be bothered by the noise when scenario 2 was performed. For cold, the reported noise disturbance was globally lower; a possible explanation could be that volunteers integrated the noise produced by the fan as wind in a snowing environment. Volunteers could less associate the noise from the heater in the chimney environment, so it was more disturbing.

Aside from this, it would have been interesting to investigate more deeply the effect of the way to produce heat. Despite no significant difference between heater at level 2 and projector, we decided to measure the temperature *a posteriori*. The thermometer was positionned at 0.5m of the device, as in our scenarios. The temperature rise is faster and higher for the projector (more than 30°C after 30s) than for the heater (24.7°C after 30s). At that time, the temperature given by the heater is stable, in opposition to the one given by the projector which is still increasing. In our scenarios, we do not allow stabilization of the projector. Regarding measured temperatures, noise disturbance, and warm sensations, the projector seems to be more efficient than the heater. This can be explained by the way to produce heat. The projector emits heat by radiation, while the heater blows warm air. Concerning cold stimulation, fan + water spray is more efficient than fan alone.

In opposite of what could be expected, our results show no significant effect of the visual cue on the thermal sensation except for the extremely cold stimulation corresponding to snow visual with a cutaneous stimulation with fan+water. This result is in opposition with Günther et al. [15] who observe an impact of the visual stimuli on the perceived temperatures when considering the neutral thermal stimulus of 32.5°C.

To compare our results with the one obtained by Günther et al. [15], we expressed all the thermic sensations (mean) on the same scale from 0 to 5 (0 corresponds to extremely cold, and 5 to extremely warm), as they were expressed on a scale from 0 to 8 in Günther et al. [15]. Although our setup is less sophisticated, it remains as efficient as "therminator" described in Günther et al. [15], as we have a device that has a same range of thermic sensation, except for the extremely cold sensation (22.5°C).

## 5 CONCLUSION AND FUTURE WORK

The generation of different cutaneous sensations in VR is a challenging task. In this work, we have described a protocol whose goals were to assess the effect of the acoustic disturbance and the impact of visual stimuli on thermal perception. We conducted two experiments. Different devices have been used to recreate cold and warm sensations. Future work will investigate the inclusion of additional multisensory cues.

**ACKNOWLEDGMENTS**

The authors would like to thank Verène Tilly Djuigui Fotso, student who started to work on the project; and Guillaume Lavoué for his fresh, experienced and benevolent outlook on the topic.